\documentclass[%
 aip,
 amsmath,amssymb,
 reprint,%
]{revtex4-1}

\usepackage{graphicx}
\usepackage{dcolumn}
\usepackage{bm}
\usepackage{float}
\usepackage[utf8]{inputenc}
\usepackage[T1]{fontenc}
\usepackage{mathptmx}
\usepackage{etoolbox}
\usepackage{xcolor}
\DeclareUnicodeCharacter{2212}{-}
\usepackage{ragged2e}  
\makeatletter
\def\@email#1#2{%
 \endgroup
 \patchcmd{\titleblock@produce}
  {\frontmatter@RRAPformat}
  {\frontmatter@RRAPformat{\produce@RRAP{*#1\href{mailto:#2}{#2}}}\frontmatter@RRAPformat}
  {}{}
}%
\makeatother
\begin{document}

\preprint{AIP/123-QED}

\title{Unlocking high hole mobility in diamond over a wide temperature range via efficient shear strain}

\author{Jianshi Sun}
\affiliation{Institute of Micro/Nano Electromechanical System and Integrated Circuit, College of Mechanical Engineering, Donghua University, Shanghai 201620, China
}%

\author{Shouhang Li}\thanks{shouhang.li@universite-paris-saclay.fr; shouhang.li@dhu.edu.cn}
\affiliation{Centre de Nanosciences et de Nanotechnologies, CNRS, Université Paris-Saclay, 10 Boulevard Thomas Gobert, Palaiseau 91120, France
}
\affiliation{Institute of Micro/Nano Electromechanical System and Integrated Circuit, College of Mechanical Engineering, Donghua University, Shanghai 201620, China
}%

\author{Cheng Shao}
\affiliation{
 Thermal Science Research Center, Shandong Institute of Advanced Technology, Jinan, Shandong 250103, China
}

\author{Zhen Tong}
\affiliation{
School of Advanced Energy, Sun Yat-Sen University, Shenzhen 518107, China
}

\author{Meng An}
\affiliation{
 Department of Mechanical Engineering, The University of Tokyo, 7-3-1 Hongo, Bunkyo, Tokyo 113-8656, Japan
}

\author{Yuhang Yao}
\affiliation{Institute of Micro/Nano Electromechanical System and Integrated Circuit, College of Mechanical Engineering, Donghua University, Shanghai 201620, China
}%

\author{Yue Hu}
\affiliation{%
CTG Wuhan Science and Technology Innovation Park, China Three Gorges Corporation, Wuhan 430010, China
}

\author{Xiongfei Zhu}
\affiliation{Institute of Micro/Nano Electromechanical System and Integrated Circuit, College of Mechanical Engineering, Donghua University, Shanghai 201620, China
}%

\author{Yifan Liu}
\affiliation{Institute of Micro/Nano Electromechanical System and Integrated Circuit, College of Mechanical Engineering, Donghua University, Shanghai 201620, China
}%

\author{Renzong Wang}
\affiliation{Institute of Micro/Nano Electromechanical System and Integrated Circuit, College of Mechanical Engineering, Donghua University, Shanghai 201620, China
}%

\author{Xiangjun Liu}\thanks{xjliu@dhu.edu.cn}
\affiliation{Institute of Micro/Nano Electromechanical System and Integrated Circuit, College of Mechanical Engineering, Donghua University, Shanghai 201620, China
}%

\author{Thomas Frauenheim}
\affiliation{School of Science, Constructor University, Bremen 28759, Germany
}
\affiliation{Institute for Advanced Study, Chengdu University, Chengdu 610106, China}

\date{\today}

\begin{abstract}
As a wide bandgap semiconductor, diamond holds both excellent electrical and thermal properties, making it highly promising in the electrical industry. However, its hole mobility is relatively low and dramatically decreases with increasing temperature, which severely limits further applications. Herein, we proposed that the hole mobility can be efficiently enhanced via slight compressive shear strain along the [100] direction, while the improvement via shear strain along the [111] direction is marginal. This impressive distinction is attributed to the deformation potential and the elastic compliance matrix. The shear strain breaks the symmetry of the crystalline structure and lifts the band degeneracy near the valence band edge, resulting in a significant suppression of interband electron-phonon scattering. Moreover, the hole mobility becomes less temperature-dependent due to the decrease of electron scatterings from high-frequency acoustic phonons. Remarkably, the in-plane hole mobility of diamond is increased by $\sim800\%$ at 800 K with a 2\% compressive shear strain along the [100] direction. The efficient shear strain strategy can be further extended to other semiconductors with face-centered cubic geometry.
\end{abstract}

\maketitle
\section{INTRODUCTION}
Enhancing carrier mobility is essential for improving the performance and reducing the power consumption of next-generation electronic devices, particularly in high-frequency and high-power applications,\cite{irds2023} where obtaining large charge carrier mobility is vital. Diamond is a rising star semiconductor owing to its ultra-wide bandgap, ultra-high thermal conductivity, and high saturation carrier velocity.\cite{wort2008diamond} However, the hole mobility of diamond is relatively lower compared to its electron mobility\cite{isberg2002high} and dramatically decreases with increasing temperature.\cite{ponce2021first} These drawbacks in hole mobility significantly constrain the potential of diamonds in CMOS,\cite{sumant2010ultrananocrystalline} high-power LEDs,\cite{koizumi2001ultraviolet} high-power lasers,\cite{williams2015efficient} and detectors.\cite{kurinsky2019diamond}

Strained silicon technology\cite{baykan2010strain,thompson2006uniaxial} is effective in promoting the hole mobility of silicon, which has been extensively utilized in industry and drives the ongoing progress of Moore’s law.\cite{lundstrom2003moore} It was confirmed that applying uniaxial compressive strain along the channel in 45 nm gate-length PMOS devices can decrease the hole effective mass\cite{yu2008first} and enhance the low-field hole mobility by 50\% and the drive current by 25\%.\cite{thompson2004logic} Although the strain engineering on the hole mobility of Si,\cite{fischetti1996band} Ge,\cite{murphy2011giant} and Ge-Sn alloy\cite{sau2007possibility} has been extensively studied, research on the diamond, which possesses a similar face-centered cubic (\textit{fcc}) structure, remains relatively scarce. This may be attributed to the extreme hardness and strength of diamond which are widely considered to render strain implementation nearly impossible.\cite{haines2001synthesis} Recently, Dang \textit{et al}.\cite{dang2021achieving} realized uniform elastic strain along the [100], [101], and [111] crystal directions by fabricating micron-nanometer scale bridge structures, thereby making it possible for the widespread application of “strained diamond” in electronic devices. It remains unclear whether strain engineering effectively promotes the hole mobility of diamond and shares a similar mechanism with other group IVA \textit{fcc} semiconductors.

Strain engineering can be classified into isotropic strain (i.e., hydrostatic strain) and shear strain.\cite{yu2005fundamentals} The isotropic strain has marginal effects on the hole mobility of \textit{fcc} geometries since it fails to break the crystal symmetry. It merely shifts energy bands without changing the valence band edge. Recently, it was found that uniaxial compressive strain is an effective way to enhance the hole mobility of wurtzite GaN due to the reverse ordering of valance bands near the valance band maximum (VBM).\cite{ponce2019route,ponce2019hole} However, a previous study shows that even a 3\%  compressive shear strain along the [111] direction cannot significantly lift the band degeneracy and has marginal effects on the hole mobility of diamond,\cite{zheng2024direct} which indicates the moderate shear strain cannot efficiently enhance the hole mobility of diamond.

In this work, it is found that the hole mobility of diamond can be dramatically increased via slight shear strain along [101] and [100] crystalline directions. Moreover, the hole mobility with shear strain becomes less temperature-dependent and maintains high values at elevated temperatures, which is of great significance in the practical applications of diamond-based electronic devices. Unlike silicon, the hole effective mass of diamond does not significantly decrease under shear strain. By employing rigorous mode-level first-principles calculations, we reveal that shear strain breaks the crystal symmetry and lifts the VBM degeneracy, thereby significantly suppressing the interband electron-phonon scattering. It is revealed that shear strain along [100] direction is most efficient to split valence bands of diamond due to the large deformation potential and the large differences in elements of the elastic compliance matrix.

\section{THEORY AND METHODS}
According to the linearized electron Boltzmann transport equation,\cite{ziman2001electrons} the mobility ($\mu$) can be expressed as
\begin{equation}
    \mu_{\alpha \beta}=\frac{e}{n_{c} \Omega} \sum_{n} \int \frac{d^{3} \mathbf{k}}{\Omega_{\mathrm{BZ}}} v_{n \mathbf{k}, \alpha} \partial_{E_{\beta}} f_{n \mathbf{k}},
    \label{SE1}
\end{equation}
where $\textit{e}$ is the elementary charge, $\alpha$ and $\beta$ are Cartesian coordinates, $n_{\textit{c}}$ is the charge carrier concentration, $\Omega$ and $\Omega_{\mathrm{BZ}}$ denote the volumes of the primitive cell and first Brillouin zone, respectively. $v_{n \mathbf{k}, \alpha}=\hbar^{-1} \partial \varepsilon_{n \mathbf{k}} / \partial k_{\alpha}$ is the electron group velocity of the electron mode with the Kohn-Sham energy $\varepsilon_{n \mathbf{k}}$, band index $n$, and wavevector $\mathbf{k}$. $\partial_{E_{\beta}} f_{n \mathbf{k}}$ is the perturbation to the Fermi-Dirac distribution due to the external electric field $\mathbf{E}$. The perturbation to the equilibrium carrier distribution $f_{n \mathbf{k}}^{0}$ (Fermi-Dirac distribution) is obtained by solving the following self-consistent equation:
\begin{equation}
\begin{aligned}
\partial_{E_{\beta}} f_{n \mathbf{k}} = & \, e \frac{\partial f_{n \mathbf{k}}^{0}}{\partial \varepsilon_{n \mathbf{k}}} v_{n \mathbf{k}, \beta} \tau_{n \mathbf{k}} \\
& + \frac{2 \pi \tau_{n \mathbf{k}}}{\hbar} \sum_{m \nu} \int \frac{d \mathbf{q}}{\Omega_{\mathrm{BZ}}} \left| g_{m n \nu}(\mathbf{k}, \mathbf{q}) \right|^{2} \\
& \times \left[ \left(n_{\mathbf{q} \nu}^{0} + 1 - f_{n \mathbf{k}}^{0}\right) \delta\left(\Delta \varepsilon_{\mathbf{k}, \mathbf{q}}^{n m} + \hbar \omega_{\mathbf{q} \nu}\right) \right. \\
& \quad + \left. \left(n_{\mathbf{q} \nu}^{0} + f_{n \mathbf{k}}^{0}\right) \delta\left(\Delta \varepsilon_{\mathbf{k}, \mathbf{q}}^{n m} - \hbar \omega_{\mathbf{q} \nu}\right)\right] \partial_{E_{\beta}} f_{m \mathbf{k} + \mathbf{q}},
\end{aligned}
\label{SE2}
\end{equation}
where $\Delta \varepsilon_{\mathbf{k}, \mathbf{q}}^{n m}=\varepsilon_{n \mathbf{k}}-\varepsilon_{m \mathbf{k}+\mathbf{q}}$, $n_{\mathbf{q} \nu}^{0}$ is the Bose-Einstein distribution and $\omega_{\mathbf{q}\nu}$ is the phonon frequency. The electron−phonon matrix element $g_{m n \nu}(\mathbf{k}, \mathbf{q})=\left(\hbar / 2 \omega_{\mathbf{q} \nu}\right)^{1 / 2}\left\langle m \mathbf{k}+\mathbf{q}\left|\Delta_{\mathbf{q} \nu} V\right| n \mathbf{k}\right\rangle$ quantifies the probability amplitude for scattering between the electronic states $n \mathbf{k}$ and $m \mathbf{k}+\mathbf{q}$, with $\Delta_{\mathbf{q} \nu} V$ the first-order differential of the Kohn−Sham potential with respect to atomic displacement. The electron-phonon scattering rate (inverse of phonon limited electron relaxation time $\tau_{n \mathbf{k}}$) is computed within the Fan-Migdal self-energy relaxation time approximation (SERTA)\cite{giustino2017electron}
\begin{equation}
\begin{aligned}
\frac{1}{\tau_{n \mathbf{k}}} = & \frac{2 \pi}{\hbar} \sum_{m \mathbf{k}+\mathbf{q}} \left| g_{m n \nu}(\mathbf{k}, \mathbf{q}) \right|^{2} \times \\
& \left\{ \left[n_{\mathbf{q} \nu}^{0} + f_{m \mathbf{k}+\mathbf{q}}^{0}\right] \delta\left(\Delta \varepsilon_{\mathbf{k}, \mathbf{q}}^{n m} + \hbar \omega_{\mathbf{q} \nu}\right) \right. \\
& \quad + \left. \left[n_{\mathbf{q} \nu}^{0} + 1 - f_{m \mathbf{k}+\mathbf{q}}^{0}\right] \delta\left(\Delta \varepsilon_{\mathbf{k}, \mathbf{q}}^{n m} - \hbar \omega_{\mathbf{q} \nu}\right) \right\}.
\end{aligned}
\label{SE3}
\end{equation}
The Dirac delta functions represent energy conservation during the scattering process, and an adaptive broadening approach\cite{ponce2021first} is employed for numerical computation. Given that diamond is a non-polar semiconductor material, only the quadrupole interactions\cite{brunin2020electron,jhalani2020piezoelectric} are included in the long-range term of electron-phonon coupling, and their effects on the hole mobility are less than 2\%. The SERTA of Eq. \ref{SE2} is employed as the initial value of $\partial_{E_{\beta}} f_{n \mathbf{k}}$, and then we iteratively solve Eqs. \ref{SE1} and \ref{SE2}, ultimately obtaining the converged value of hole mobility.

The Quantum Espresso package\cite{giannozzi2009quantum} is employed for the first-principles calculations with the Perdew−Burke−Ernzerhof (PBE) form of the exchange−correlation functional\cite{perdew2008restoring} and optimized full relativistic norm-conserving pseudopotentials\cite{hamann2013optimized} from PseudoDojo.\cite{van2018pseudodojo} When strain is applied, the primitive cell is uniformly deformed along the strained direction to the desired level, while full relaxation is permitted in the other orthogonal directions. The spin-orbit coupling (SOC) is included in the electronic band structure calculations. The accuracy of the electron band structure from PBE exchange−correlation functional is validated by comparing it with that from the screened Heyd-Scuseria-Ernzerhof (HSE) hybrid functional,\cite{heyd2003hybrid} as shown in [Figure S2, Supplementary Material]. The harmonic force constants are calculated from density-functional perturbation theory (DFPT).\cite{fugallo2013ab,baroni2001phonons} The dynamical quadrupole tensors are calculated from linear response theory as implemented in ABINIT.\cite{gonze2020abinit,romero2020abinit} The in-house modified EPW package\cite{ponce2016epw} is employed to make the calculations on dense \textbf{k}/\textbf{q} meshes feasible. Additional computational details and convergence tests are presented in Sections I-V of the Supplementary Material.

\begin{figure*}[hbpt]
    \includegraphics[width=2\columnwidth]{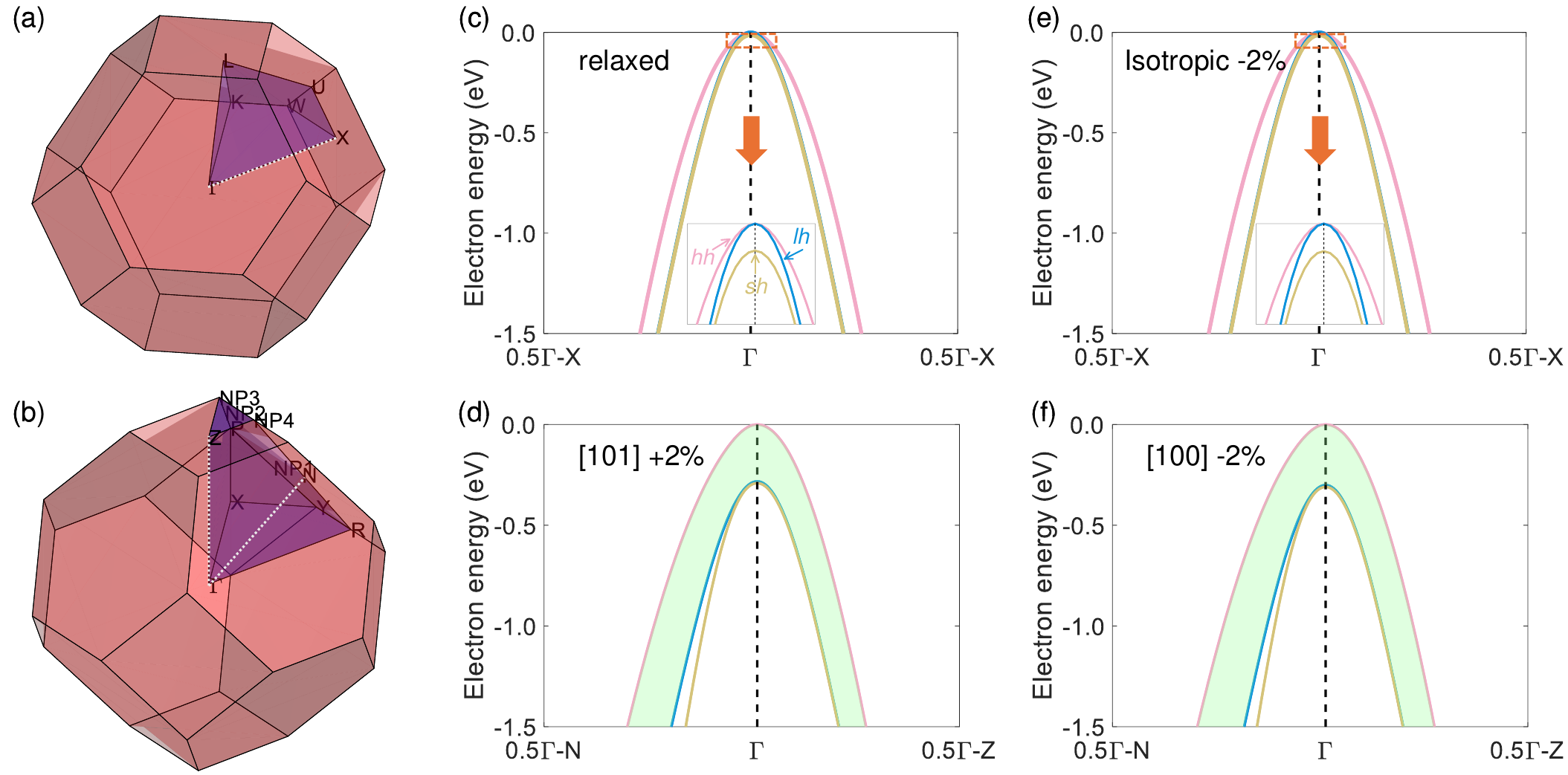}
    \caption{\justifying The Brillouin zone and the irreducible wedge (shadow region) for diamond crystalline geometries with (a) space group $F d \overline{3} m$ and (b) space group $I 4_{1} / \textit{amd}$. Valence bands of (c) relaxed structure, (d) the structure with +2\% shear strain along the [101] direction, (e) the structure with -2\% isotropic strain, and (f) the structure with -2\% shear strain along the [100] direction. The high-symmetry paths are marked by white dashed lines in the BZ zone. The electron energy is normalized to the VBM. Zoomed-in band structures near the VBM are inserted at the bottom of (c) and (e). The light green ribbon represents significant band splitting induced by shear strain, which notably suppresses interband electron-phonon scattering.}
    \label{fig: Figure 1}
\end{figure*}

\begin{figure*}[hbpt]
    \includegraphics[width=2\columnwidth]{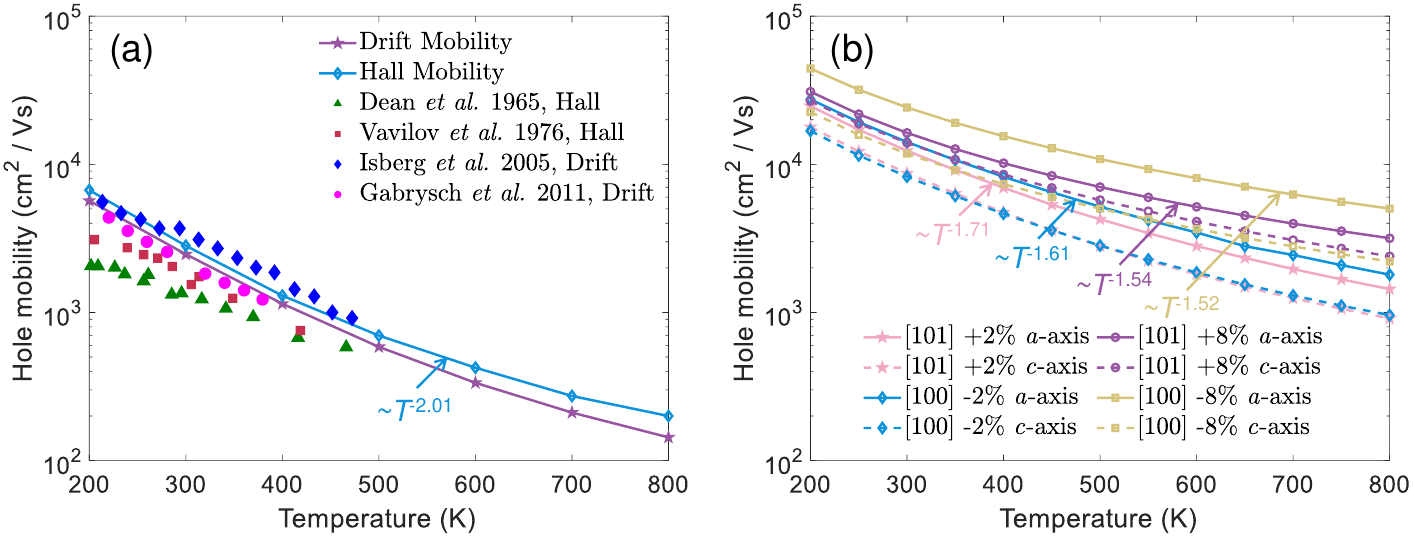}
    \caption{\justifying (a) Drift mobility and Hall mobility as a function of temperature for relaxed diamond. The scatters are experimental data reported by Dean \textit{et al.}\cite{dean1965intrinsic} (triangle), Vavilov \textit{et al.}\cite{vavilov1976semiconducting} (square), Isberg \textit{et al.}\cite{isberg2005temperature} (diamond), and Gabrysch \textit{et al.}\cite{gabrysch2011electron} (circle), respectively. (b) In-plane (\textit{a}-axis) and out-of-plane (\textit{c}-axis) Hall mobilities for shear strains along the [101] and [100] directions. }
    \label{fig: Figure 2}
\end{figure*}

\section{RESULTS AND DISCUSSIONS}
The crystalline geometry of the relaxed diamond is \textit{fcc} (space group $F d \overline{3} m$, no. 227) and its first Brillouin zone (BZ) is a truncated octahedron, as shown in Fig. \ref{fig: Figure 1}(a). The isotropic strain does not alter the crystal symmetry of the diamond, thereby there is almost no change in the profile of valence bands near the VBM, as presented in 
Figs. \ref{fig: Figure 1}(c) and \ref{fig: Figure 1}(e). It should be noticed that the SOC removes the triple degeneracy at the VBM, resulting in the double degeneracy of the heavy-hole (\textit{hh}) and light-hole (\textit{lh}) bands, while the split-off hole (\textit{sh}) band lies 13.4 meV and 13.7 meV lower in energy, as shown in the insets of Figs. \ref{fig: Figure 1}(c) and \ref{fig: Figure 1}(e). The clustered phenomenon of valance bands is common in semiconductors, and it induces a lot of electron-phonon scatterings.\cite{ponce2019route,ponce2019hole,sun2024giant} In contrast, the crystal symmetry is broken with the slight shear strain applied to the \textit{fcc} diamond. The space group is changed to $I 4_{1} / \textit{amd}$ (no. 141) and the first BZ is shown in Fig. \ref{fig: Figure 1}(b). As a result, the double degeneracy of the \textit{hh} and \textit{lh} bands is lifted under a 2\% shear strain along both [101] and [100] directions, arousing a large energy splitting of $\sim0.3$\ eV (highlighted by light green ribbons in Figs. \ref{fig: Figure 1}(d) and \ref{fig: Figure 1}(f). As the shear strain increases, the gap between the \textit{hh} band and the group of \textit{lh} and \textit{sh} bands is broadened [Figure S6, Supplementary Material]. This alteration in valence band structure significantly suppresses the interband electron-phonon scattering, which will be discussed later. Furthermore, the hole effective mass of \textit{hh} band near the VBM is almost unchanged with shear strain (see Table S1, Supplementary Material), which contrasts with the observation in silicon.\cite{yu2008first,roisin2024phonon}

 \begin{figure}[hbpt]
    \centering
    \includegraphics[width=0.95\columnwidth]{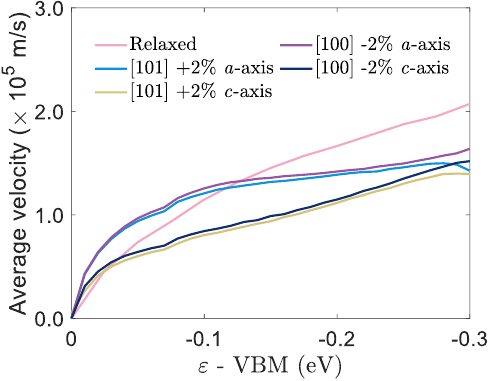}
    \caption{The in-plane (\textit{a}-axis) and out-of-plane (\textit{c}-axis) average electron group velocity for diamonds without/with shear strains. The average electron group velocity is calculated according to the equation $\bar{v}^{\alpha}(\varepsilon)=\sum_{n\mathbf{k}}\left|v^{\alpha}_{n\mathbf{k}}\right| \delta\left(\varepsilon-\varepsilon_{n\mathbf{k}}\right) / \sum_{n\mathbf{k}} \delta\left(\varepsilon-\varepsilon_{n\mathbf{k}}\right)$.}
    \label{fig: Figure 3}
\end{figure}

There is little band degeneracy lifting with -2\% shear strain along the [111] direction, in contrast to the same shear strain along the [100] direction [Figure S7, Supplementary Material]. It should be noted that the shear strain along the [111] direction is not equivalent to the isotropic strain. The off-diagonal elements of the shear strain tensor are zero under isotropic strain, while they are non-zero when the shear strain is applied along the [111] direction. The splitting energy ($\Delta$) between \textit{hh} and \textit{lh} bands under [100] and [111] shear strains can be expressed as\cite{hensel1963cyclotron}
\begin{equation}
\begin{aligned}
\Delta_{100}=\frac{4}{3} \epsilon D_{u}\left(s_{11}-s_{12}\right),\\
\Delta_{111}=\frac{4}{3} \epsilon D_{u^{\prime}}\left(s_{44} / 2\right),
\end{aligned}
\label{SE4}
\end{equation}
where $\epsilon$ is the strain applied along the crystallographic directions. $D_{u}$ ($D_{u^{\prime}}$) is the Kleiner-Roth valence band deformation potential along the [100] ([111]) direction under compressive shear strain. $s_{i j}$ are the elements of the elastic compliance matrix. These parameters for {\textit{fcc}} semiconductors are listed in Table \ref{tab:Table1}. It is found that diamond has a ratio of $\Delta_{100}$ to $\Delta_{111}$ ($\sim 9$), which is consistent with the ratio of the splitting energies between \textit{hh} and \textit{lh} bands under [100] and [111] shear strains from first-principles calculation. The impressive magnitude of $\Delta_{100}$ is attributed to the large values of $D_{u}$ and $s_{11} - s_{12}$, while the small value of $\Delta_{111}$ is primarily due to the low value of $s_{44}$. In contrast, silicon holds a much smaller $\Delta_{100}$ due to the much smaller $D_{u}$, which offsets the large value of $s_{11}$ - $s_{12}$. We can further obtain the splitting energy for another compound semiconductor 3C-SiC based on Eq. \eqref{SE4} and it is consistent with that from first-principles calculation (Figure S8, Supplementary Material). 
 
\begin{table*}[hbpt]
    \centering
    \caption{Deformation potential parameters ($b/d$),\cite{blacha1984deformation,lambrecht1991calculated} elements of the elastic stiffness matrix ($c_{i j}$),\cite{kittel2018introduction} and the ratio of the 
    splitting energies of valence bands. The Kleiner-Roth deformation potentials\cite{kleiner1959deformation} are related to Pikus and Bir's notation\cite{bir1974symetry} by $D_{u} = -\frac{3}{2} \textit{b}$ \text{and} $D_{u'} = -\frac{1}{2} \sqrt{3} \textit{d}$. Elastic stiffness constants and elastic compliance constants for cubic crystals are related by $c_{44}=(s_{44})^{-1}$ and $c_{11}-c_{12}=\left(s_{11}-s_{12}\right)^{-1}$.}
    \label{tab:Table1}
    \renewcommand{\arraystretch}{1.5} 
    \setlength{\tabcolsep}{10pt} 
    \begin{tabular}{ccccccc} 
        \hline
        Material & \textit{b}/(eV) & \textit{d}/(eV) & $\textit{c}_{11}/(10^{11} \, \text{N/m}^2$) & $\textit{c}_{12}/(10^{11} \, \text{N/m}^2$) & $\textit{c}_{44}/(10^{11} \, \text{N/m}^2$) \ & $\Delta_{100}$/$\Delta_{111}$\\
        \hline
        Diamond & -8.1 & -7.5 & 10.76 & 1.25 & 5.76 & 9.1 \\
        \hline
        Silicon & -2.2 & -5.0 & 1.66 & 0.64 & 0.80 & 4.7  \\
        \hline
        3C-Silicon Carbide & -2.2 & -6.3 & 3.52 & 1.20 & 2.33 & 4.9  \\
        \hline
    \end{tabular}
\end{table*}
 
Fig. \ref{fig: Figure 2}(a) shows the phonon-limiting hole drift mobility and Hall mobility of the relaxed diamond, as well as existing experimental data. Overall, the theoretically predicted results agree well with experimental data in a wide temperature range. The quantitative differences between our work and Refs.\cite{dean1965intrinsic,vavilov1976semiconducting,isberg2005temperature,gabrysch2011electron} may stem from the outdated nature of the existing experimental data, highlighting the need for more recent measurements for meaningful comparison. It should be noted that achieving a moderate carrier concentration in diamond typically requires a very high doping level ($\sim$ $10^{20} \, \mathrm{cm}^{-3}$) due to the high activation energy and the resulting limited activation efficiency.\cite{brillas2011synthetic} Therefore, a substantial amount of hole dopants is required to facilitate \textit{p}-type carrier conduction, where the roles of ionized and neutral impurities scattering are crucial. These aspects will be discussed in our companion paper. The drift mobility is quite close to the Hall mobility, indicating that the Hall factor is close to one. Given that most experimental data are Hall mobility, the following discussion will focus on the Hall mobility of diamond.

\begin{figure*}[hbpt]
    \includegraphics[width=2\columnwidth]{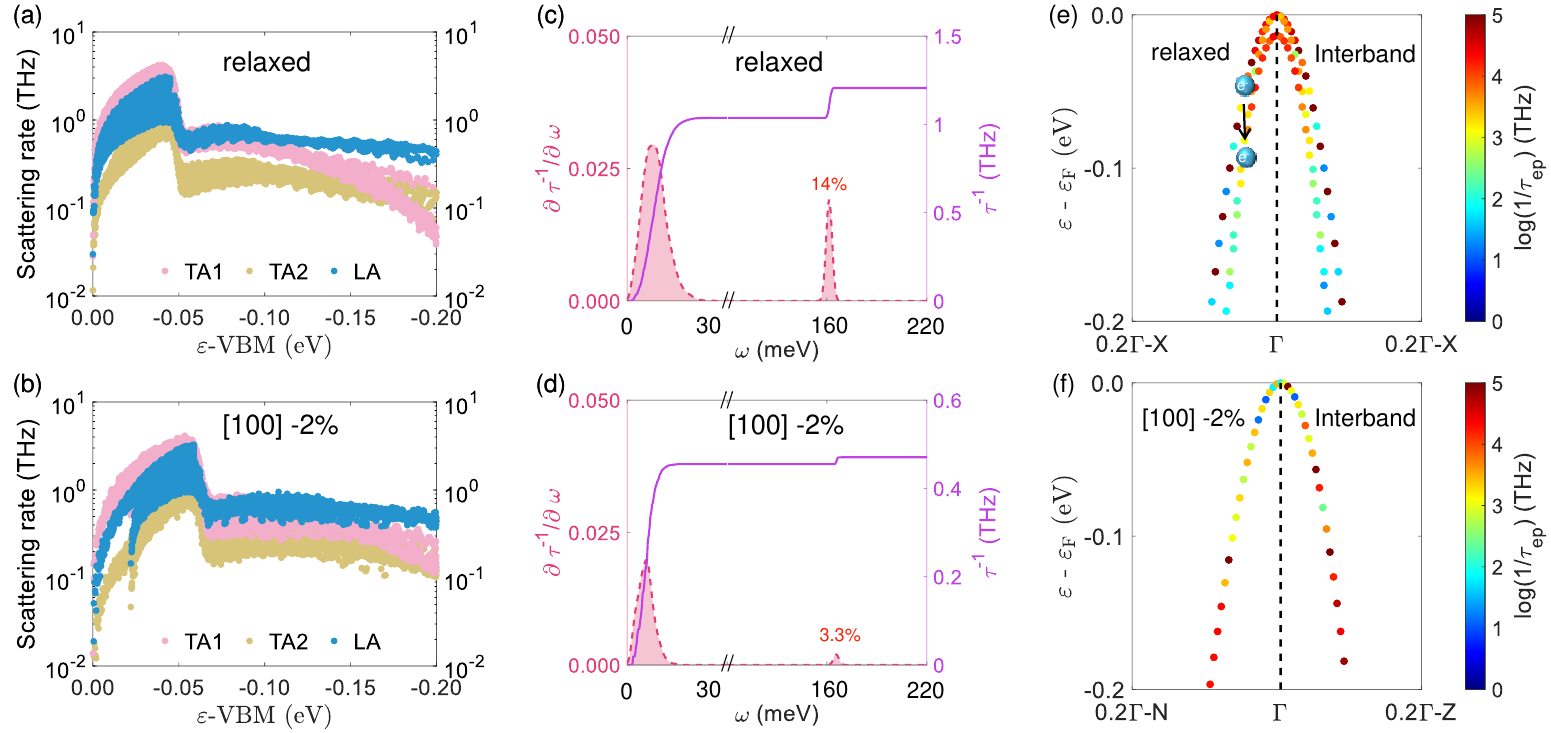}
    \caption{\justifying Mode-resolved room-temperature hole scattering rates of diamonds (a) without and (b) with -2\% shear strain. Spectral decomposed hole scattering rates for (c) relaxed and (d) -2\% shear strain as a function of phonon energy calculated at 39 meV away from the valence band edges. The peaks indicate $\partial \tau^{-1} / \partial \omega$ (left axis), and the purple solid line represents the cumulative integral $\tau^{-1}$ (right axis). The percentages indicate the cumulative contribution of optical phonons to electron-phonon scattering. Projection of interband electron-phonon scattering rates onto the band structure for (e) relaxed and (f) -2\% shear-strained diamond. }
    \label{fig: Figure 4}
\end{figure*}

The room-temperature hole mobility of the relaxed diamond is 2820 $\mathrm{cm\textsuperscript{2}/(Vs)}$. There is a dramatic enhancement in hole mobility when shear strain is applied, as shown in Fig. \ref{fig: Figure 2}(b). Specifically, the in-plane mobility is increased to 12355 $\mathrm{cm\textsuperscript{2}/(Vs)}$ under 2\% tensile (+2\%) shear strain along the [101] direction, and it reaches 14148 $\mathrm{cm\textsuperscript{2}/(Vs)}$ from the same initial value under 2\% compressive (-2\%) shear strain along the [100] direction. Correspondingly, the out-of-plane mobility reaches 8715 $\mathrm{cm\textsuperscript{2}/(Vs)}$ for the tensile shear strain case and 8252 $\mathrm{cm\textsuperscript{2}/(Vs)}$ for the compressive shear strain case. It is more efficient to increase the hole mobility when the compressive shear strain is applied to the [100] direction due to the slightly larger splitting between the \textit{hh} band and the group of \textit{lh} and \textit{sh} bands (see Fig. \ref{fig: Figure 1}). As the shear strain increases to 8\%, the enhancement in hole mobility gradually approaches saturation. The in-plane hole mobility reaches 24190 $\mathrm{cm\textsuperscript{2}/(Vs)}$, which is one order of magnitude larger than that of relaxed diamond and is the highest recorded value in bulk semiconductors. Details regarding the impact of 5\% shear strain on hole mobility are provided in the Supplementary Material. The hole mobility typically decreases exponentially with temperature ($\mu \sim T^{-n}$) since phonons are significantly activated, which degrades the performance of electronic devices. For the relaxed case, the value of exponent \textit{n} in the temperature dependence is -2.01, which is in excellent agreement with the experimental value of -2.06.\cite{gabrysch2011electron} We show that hole mobility becomes less temperature-dependent for diamonds with shear strains. The exponent \textit{n} values of in-plane hole mobility are located in the range of -1.52 to -1.71.

The hole mobility is mainly determined by the electron group velocity and relaxation time. The average electron group velocity for diamonds without/with shear strains are shown in Fig. \ref{fig: Figure 3}. It is found that the in-plane electron group velocity for electron states near valence band edge is relatively higher than that of the relaxed diamond, whereas the out-of-plane electron group velocity exhibits the opposite trend. The anisotropy in the electron group velocity results in the anisotropic hole mobility shown in Fig. \ref{fig: Figure 2}(b). Overall, the influences of shear strains on the electron velocity are marginal, which is also reflected in the band structures shown in Fig. \ref{fig: Figure 1}. The slopes of valance bands for diamonds without/with shear strains are quite similar. The phenomenon is different from the observation in hexagonal wide bandgap semiconductors, where the out-of-plane electron group velocity is significantly increased with uniaxial compressive strain.\cite{ponce2019route,ponce2019hole,sun2024giant} Therefore, the electron velocity cannot explain the dramatic increase in hole mobility presented in Fig. \ref{fig: Figure 2}(b).

To understand the increase in the hole mobility induced by shear strains, the mode-resolved hole scattering rates for diamonds without/with strain are shown in Figs. \ref{fig: Figure 4}(a) and \ref{fig: Figure 4}(b), respectively. We mainly focus on the electron modes with energies from VBM to 50 meV below VBM since those modes have significant effects on electron transport. Within this range, the influence of optical phonons on hole mobility is negligible due to the energy conservation is not adequately satisfied during the electron-phonon scattering process. Therefore, hole transport is mainly scattered by acoustic phonons. For the relaxed case, holes are primarily scattered by longitudinal acoustic (LA) and first transverse acoustic (TA1) phonons. There is a significant reduction in the scattering rates regarding LA and TA1 phonons for the electron modes close to the VBM when the shear strain is applied, thereby increasing the hole mobility. This can be further validated by the spectral decomposed angular-averaged scattering rate at a phonon energy of 3\textit{$k_b$}\textit{T}/2 (39 meV) away from the VBM, as shown in Figs. \ref{fig: Figure 4}(c) and \ref{fig: Figure 4}(d). It is found that the main scattering channel comes from acoustic phonons for both relaxed and strained diamonds. The peak associated with the phonon energy of 160 meV (the lowest value of optical phonon frequency) can be attributed to the availability of optical phonon absorption.

We further project the interband electron-phonon scattering rates onto the band structure for diamonds without/with strain, as shown in Figs. \ref{fig: Figure 4}(e) and \ref{fig: Figure 4}(f). The intraband electron-phonon scattering rates are shown in the [Figure S13, Supplementary Material]. Here, we only present the \textit{hh} band for the strained case, as the other bands have negligible contributions to hole mobility. The interband electron-phonon scattering rates are significantly larger than the intraband electron-phonon scattering rates in the relaxed case due to the clustered behavior of valance bands. In contrast, there is a significant reduction in the interband electron-phonon scattering rates for the shear-strained diamond, attributed to the substantial energy splitting in the valence bands, as discussed before. This energy splitting prohibits energy conservation and diminishes the available electron-phonon scattering channels. Moreover, intraband electron-phonon scattering processes are primarily mediated by low-frequency acoustic phonons, while the interband scattering processes are mainly assisted by high-frequency acoustic phonons. This mechanism can be detected from Figs. \ref{fig: Figure 4}(c) and \ref{fig: Figure 4}(d), where the frequencies of acoustic phonons limiting hole transport become lower. The relative increase in the number of low-frequency acoustic phonons with temperature is not as significant as that of high-frequency acoustic phonons. Therefore, the hole mobility for strained diamonds becomes less temperature-dependent, as shown in Fig. \ref{fig: Figure 2}(b).

Finally, we discuss practical methods for achieving \textit{p}-type strained diamond. Experimentally, \textit{p}-type doping of diamond can be efficiently achieved using boron delta doping techniques.\cite{geis2018progress} Furthermore, the shear strain can be achieved in heteroepitaxial-grown diamonds. Iridium is considered the most promising substrate for diamond heteroepitaxy and a large lattice mismatch of $\sim7$\% can be realized\cite{lee2016epitaxy}. It should be noted that diamond epitaxial layers are typically deposited on thick films, which can lead to strain relaxation through mismatch dislocations and cracks.\cite{tang2016stress} To prevent mismatch dislocations from impeding improvements in hole mobility, it is essential to grow appropriately strained films by maintaining the film thickness below the critical value.\cite{fischer1994new} Moreover, strain-induced alterations in impurity energies can effectively enhance the doping solubility in various semiconductor materials, establishing a synergistic relationship between strain and doping.\cite{zhu2010strain}

\section{CONCLUSIONS}
In summary, we demonstrated that the shear strain along the [101] and [100] directions is an efficient strategy to enhance the hole mobility of diamond, while the shear strain along the [111] direction is inefficient. Unlike silicon, the hole effective mass of diamond is not significantly decreased by shear strains. Instead, the shear strain breaks the symmetry of the diamond crystalline structure, effectively lifting the band degeneracy near the valence band maximum. Shear strain along the [100] direction is most efficient to split the valence bands of diamond, due to the large deformation potential and significant differences in the elements of the elastic compliance matrix. As a result, the interband electron-phonon scattering mediated by acoustic phonons is significantly suppressed and the hole mobility can be increased by one order of magnitude under 8\% shear strain. Moreover, the hole mobility of diamond with shear strains becomes less temperature-dependent due to the reduction of electron scatterings from high-frequency acoustic phonons. This is of great importance for the application of power electronics, where the self-heating effect is pronounced, and maintaining high mobility at elevated temperatures is highly desirable.

\section{SUPPLEMENTARY MATERIAL}
Detailed information regarding computational details, the validation of Wannierization, PBE and HSE electronic band structures, convergence tests on the drift and Hall mobilities, carrier concentration impact on hole mobility, band structure with and without spin−orbit coupling, band structures of diamond and 3C-SiC via shear strain, the impact of shear strain on hole mobility, average electron group velocity, mode-resolved and spectral decomposed hole scattering rates with shear strain, intraband electron-phonon scattering rates, hole effective masses without and with shear strain.

\section{ACKNOWLEDGMENTS}
We would like to thank Dr. Ao Wang from Université catholique de Louvain for valuable discussions. S.L. was supported by the National Natural Science Foundation of China (Grant No. 12304039), the Shanghai Municipal Natural Science Foundation (Grant No. 22YF1400100), and the Fundamental Research Funds for the Central Universities (Grant No. 2232022D-22). J.S. was supported by the Fundamental Research Funds for the Central Universities (Grant No. CUSF-DH-T-2024061). X.L. was supported by the Shanghai Municipal Natural Science Foundation (Grant No. 21TS1401500) and the National Natural Science Foundation of China (Grants Nos. 52150610495 and 12374027). The computational resources utilized in this research were provided by National Supercomputing Center in Shenzhen.

\section{DATA AVAILABILITY}
The data that support the findings of this study are available from the corresponding author upon reasonable request.

\section{REFERENCES}
\bibliography{aipsamp}

\end{document}